%% file: paper.tex
\documentclass[sigconf]{acmart}
\def\noeditingmarks{}
\usepackage{balance}

\input{preamble}
\begin{document}

\title{What's a Little Leakage Between Friends?}
\author{Sebastian Angel}
\affiliation{University of Pennsylvania}

\author{David Lazar}
\affiliation{MIT CSAIL}

\author{Ioanna Tzialla}
\affiliation{New York University
\vspace*{1em}
}

\copyrightyear{2018} 
\acmYear{2018} 
\setcopyright{acmlicensed}
\acmConference[WPES'18]{2018 Workshop on Privacy in the Electronic
Society}{October 15, 2018}{Toronto, ON, Canada}
\acmBooktitle{2018 Workshop on Privacy in the Electronic Society
(WPES'18), October 15, 2018, Toronto, ON, Canada}
\acmPrice{15.00}
\acmDOI{10.1145/3267323.3268958}
\acmISBN{978-1-4503-5989-4/18/10}
\fancyhead{}
%\settopmatter{printccs=false, printacmref=false}

\input{abs}

\date{}
\maketitle

\input{intro}
\input{bg}
\input{problem}
\input{solutions}
\input{attack}
\input{defense}
\input{disc}
\input{acks}

\frenchspacing

\begin{flushleft}
\setlength{\parskip}{0pt}
\setlength{\itemsep}{0pt}
\bibliographystyle{abbrv}
\balance
\bibliography{conferences,paper} 
\end{flushleft}
\end{document}

%% file: preamble.tex
\usepackage{booktabs}  % for tables
\usepackage{dcolumn}   % for tables
\usepackage{multirow}  % for tables
\usepackage{rotating}
\usepackage{xspace}
\usepackage{hyphenat}  % supplies \hyp{}, which tells tex that it can 
\usepackage{enumerate}

\usepackage{wrapfig}
\usepackage{textcomp}
\usepackage{tabularx}
\usepackage{pifont}

\usepackage{wrapfig}

\usepackage{ulem}
\usepackage{colortbl} % colorful columns and rows
\usepackage{array}    % needed for 'b' argument in tabular preamble

\usepackage{dblfloatfix}

\usepackage{algpseudocode}
\algrenewcomment[1]{\hfill// #1}%
\algnotext{EndFunction}
\algnotext{EndFor}
\algnotext{EndIf}

\usepackage{amsmath,amscd}
\usepackage{amssymb}
\usepackage{amsfonts}
\usepackage{amsthm}

\algrenewcommand\algorithmicindent{1em}
\algrenewcommand{\algorithmiccomment}[1]{{\color{CadetBlue}\hfill// #1}}

\usepackage[noeka]{mathrmletter}

\newif\ifextended
\newif\iflongbatching
\newif\ifsubmission
\newif\ifelementary

\ifx\buildextended\undefined
\else
    \extendedtrue
\fi

\ifx\noeditingmarks\undefined
   \newcommand{\pgwrapper}[3]{\begingroup \color{#1} #2: #3 \endgroup}
   \newcommand{\pgwrapperb}[1]{\textbf{#1}}
\else
   \newcommand{\pgwrapperb}[1]{}
   \newcommand{\pgwrapper}[3]{}
\fi

\def\hn{\usefont{OT1}{phv}{mc}{n}\selectfont}

\newcommand{\mpfont}{\hn\scriptsize}

\ifx\noeditingmarks\undefined
    \newcommand{\MPworker}[2]{{\color{#1}\vrule\vrule}{\marginpar{\color{#1}\mpfont #2}}}
\else
    \newcommand{\MPworker}[2]{}
\fi

\ifx\noeditingmarks\undefined
    
\else
    
\fi

\newcommand{\Att}{\mathcal{A}}

\theoremstyle{definition}

\newtheorem{definition}{Definition}

\makeatletter
\renewcommand*{\@fnsymbol}[1]{\ensuremath{\ifcase#1\or \star\or \dagger\or \ddagger\or
   \mathsection\or \mathparagraph\or \|\or **\or \dagger\dagger
   \or \ddagger\ddagger \else\@ctrerr\fi}}
\makeatother

\makeatletter
\def\imod#1{\allowbreak\mkern10mu({\operator@font mod}\,\,#1)}
\makeatother


\def\compactify{\itemsep=0in \topsep=2pt \parsep=0.00in \partopsep=0pt
\leftmargin=2em}
\let\latexusecounter=\usecounter

\newenvironment{myitemize}%
  {\begin{list}{\labelitemi}{\itemsep3pt \topsep3pt \parsep0.00in
  \partopsep=3pt \leftmargin1.2em}}%
  {\end{list}}
  {\begin{list}{\labelitemi}{\itemsep1pt \topsep2pt \parsep0.00in
  \partopsep=0pt \leftmargin1.2em}}%
  {\end{list}}
  {\begin{list}{\labelitemi}{\itemsep2pt \topsep2pt \parsep0.00in
  \partopsep=0pt \leftmargin1.2em}}%
  {\end{list}}
  {\begin{list}{\threequartdash}{\itemsep3pt \topsep3pt \parsep0.00in
  \partopsep=3pt \leftmargin1.5em}}%
  {\end{list}}

\newenvironment{myenumerate}
  {\def\usecounter{\compactify\latexusecounter}
   \begin{enumerate}}
  {\end{enumerate}\let\usecounter=\latexusecounter}

\def\compactsortof{\itemsep=0in \topsep=2pt \parsep=0.00in \partopsep=0pt
\leftmargin=1.7em}

\def\compactsqueeze{\itemsep=0pt \topsep0pt \parsep=0ex \partopsep=0pt
\leftmargin=1.63em}

\ifx\normalpar\undefined
  
\else
  
\fi

\def\discretionaryslash{\discretionary{/}{}{/}}
{\catcode`\/\active
\gdef\URLprepare{\catcode`\/\active\let/\discretionaryslash
        \def~{\char`\~}}}%
\def\URL{\bgroup\URLprepare\realURL}%
\def\realURL#1{\tt #1\egroup}%

%% file: abs.tex
\begin{abstract}
This paper introduces a new attack on recent messaging systems that protect 
  communication metadata.
The main observation is that if an adversary manages to compromise a user's
  friend, it can use this compromised friend to learn information
  about the user's \emph{other} ongoing conversations.
Specifically, the adversary learns whether a user is sending other messages or
  not, which opens the door to existing intersection and disclosure attacks.
To formalize this \emph{compromised friend attack}, we present an abstract 
  scenario called the \emph{exclusive call center problem} that 
  captures the attack's root cause, and demonstrates that it is 
  independent of the particular design or
  implementation of existing metadata-private messaging systems.
We then introduce a new primitive called a \emph{private answering machine} 
  that can prevent the attack.
Unfortunately, building a secure and efficient instance of this primitive 
  under only computational hardness assumptions does not appear possible.
Instead, we give a construction under the assumption that users 
  can bound their maximum number of friends and are okay leaking 
  this information.
\end{abstract}

%% file: intro.tex
\section{Introduction}\label{s:intro}

In the past few years there has been a renaissance of messaging
  systems~\cite{angel16unobservable, vandenhoof15vuvuzela, lazar16alpenhorn, 
  kwon17atom, tyagi17stadium, alexopoulos17mcmix} that allow users to 
  communicate online without their messages being observed by ISPs, companies, 
  or governments.
These systems target a property called metadata-privacy which is
  stronger than end-to-end encryption: encryption hides the content of 
  messages but it does not hide their existence nor any of their associated 
  metadata (identity of the sender or recipient, frequency, time,
  and duration of communication, etc.).
While hiding metadata has been the subject of a long line of work 
  dating back three decades~\cite{chaum81untraceable}, there is renewed 
  interest due to a proliferation of controversial surveillance 
  practices~\cite{guardian13nsa, intercept14new, guardian15gchq,
  eff15dragnet, tor15cmu}, and the monetization of users' private 
  information~\cite{wired12shady, cnet12facebook, forbes13att, reuters16yahoo}.

Existing metadata-private messaging (MPM) systems guarantee that as 
  long as the sender and the recipient of a message are not compromised, their 
  communication cannot be observed by an adversary (the adversary 
  learns that users are part of the system, but not whether they communicate).
If either the sender or the recipient is compromised, MPM systems
  provide no guarantees (e.g., a compromised sender could trivially 
  disclose to whom it is sending a message).
In this paper we investigate whether an adversary---by 
  compromising and leveraging a user's friends---can learn anything about 
  the user's \emph{other} ongoing communications.

At first glance the answer to the above question appears to be no (assuming 
  that the user does not voluntarily disclose the existence of other 
  communications to compromised friends).
After all, the guarantees of MPM systems 
  should prevent the adversary from learning anything about 
  conversations between uncompromised clients.
Nevertheless, we find that this is not actually the case: engaging in 
  a conversation with a compromised client consumes a limited resource, namely
  the number of concurrent conversations that a user can support.
By observing a client's responses (or lack thereof), a compromised friend can 
  learn whether the user has fully utilized this limited resource (i.e., the 
  user is busy talking to others).
In Section~\ref{s:attack} we show how this one bit of information enables 
  existing intersection and disclosure attacks~\cite{raymond00traffic, agrawal03disclosure, 
  kesdogan04hitting, kesdogan09breaking, danezis03statistical, mallesh10reverse, 
  troncoso08perfect,danezis09vida, danezis04statistical, danezis07two, 
  perezgonzales12understanding} that
  invalidate MPM systems' guarantees.

More interestingly, our \emph{compromised friend attack} applies to all 
  MPM systems that support a notion of 
  \emph{dialing}~\cite{lazar16alpenhorn} (or any 
  other mechanism that allows clients to start new conversations over time).
We give a formal characterization of the attack with a scenario that
  we call the \emph{exclusive call center problem}, which abstracts away
  the design or implementation of MPM systems.
We then introduce a primitive called a \emph{private answering
  machine} that solves the abstract problem and can be used
  by clients of MPM systems to prevent the compromised friend attack.
In particular, clients use a private answering machine 
  to select with which friends to communicate, while guaranteeing that 
  compromised friends learn no information about other ongoing communications.

Unfortunately, building a cryptographically-secure private 
  answering machine that does not require placing assumptions on the
  number of callers (i.e., the number of friends that a user can have) or 
  incurring prohibitive delay or bandwidth appears hard.
We compromise on this point and give a construction that 
  can be used by MPM systems under the assumption that users can place a
  bound on their maximum number of friends.
Our construction has two limitations: (1) it leaks the bound chosen by the 
  user, and (2) it increases the latency of communication between a 
  pair of users proportional to the chosen bound.
Despite these limitations, our work addresses a previously overlooked attack 
  and allows users in MPM systems to communicate without 
  leaking sensitive information. 

In summary, the contributions of this work are:
\begin{myitemize}

\item An abstraction that captures the leakage 
  from oversubscribing a fixed resource in the presence of
  adversarial probing~(\S\ref{s:problem}). 

\item The \emph{compromised friend attack} which exploits the fixed 
  communication capacity of MPM systems~(\S\ref{s:attack}).

\item The construction of a private answering machine~(\S\ref{s:solutions:bounded}) 
  that can be used in MPM systems to avoid leaking information
  to compromised friends about users' other ongoing conversations~(\S\ref{s:defense}).

\end{myitemize}

%% file: bg.tex
\section{Background}%
\label{s:bg}

The goal of metadata-private messaging systems~\cite{vandenhoof15vuvuzela,
  kwon17atom, alexopoulos17mcmix, tyagi17stadium, angel16unobservable}
  is to allow a pair (or group) of \emph{friends} to exchange bidirectional 
  messages without leaking metadata to any party besides the sender and the 
  recipient.
A pair of users are friends if they have previously shared a secret, either
  out-of-band (e.g., in person at a coffee shop), or in-band with 
  an \emph{add friend} protocol~\cite{lazar16alpenhorn}.
Users in these systems exchange a fixed number of messages with their friends 
  in discrete time epochs called rounds; users participate in every round even 
  if they are idle.
This ensures that an attacker that monitors the network cannot tell when users
  are actively communicating with their friends or starting/stopping 
  conversations.
This also places a bound on the number of active conversations that a user can 
  have at any time; we refer to this as the client's 
  \emph{communication capacity}.

Once a client reaches its communication capacity, it cannot send 
  messages to other friends until it ends an existing conversation.
As a result, clients use a separate \emph{dialing protocol} to 
  coordinate the start and end of conversations.
In a dialing protocol, a client sends a short message (a few bits) 
  to a friend regardless of whether the friend's client has reached its communication 
  capacity.
The dialing message is sufficient to notify a user that one of their friends 
  wishes to communicate, and to agree on a round to start the 
  conversation~\cite{lazar16alpenhorn}.
There are multiple ways in which a client can react to a dialing message.
Some natural choices are:

\begin{myitemize}
\item If the client has not reached its communication capacity, it can 
  automatically accept the call and start a new conversation.
\item The client could prompt the user (similar to calling a friend in Skype), 
  who can choose to accept or reject the call.
\item If at capacity, the client could randomly end an existing conversation 
  to make room for a new one.
\end{myitemize}

Each of these choices is problematic. 
If the client's communication capacity is $1$ (as in some of the existing
  systems~\cite{vandenhoof15vuvuzela, tyagi17stadium}) and the client 
  automatically accepts calls, then any of the client's friends can easily 
  learn when the client is \emph{not} active in a conversation simply by 
  calling.
Leaving the choice to the users is slightly better since the user can choose to 
  ignore or delay accepting some calls, but their choices can still 
  inadvertently lead to intersection attacks.
Ending conversations randomly hurts usability and might still leak information.
The goal of the next section is to formalize the desired properties of 
  the client's answering mechanism.

%% file: problem.tex
\section{Exclusive call center problem}%
\label{s:problem}

In order to avoid the details of particular MPM systems, we 
  introduce an abstract scenario called the \emph{exclusive call center problem.}
It consists of a call center that has $k$ operators capable
  of receiving calls (i.e., the call center has
  communication capacity $k$).
The call center promises exclusivity to a single organization.
This might be desirable to ensure high quality of service, 
  for legal reasons, or to prevent the accidental leak of trade or business 
  secrets to callers of a different organization.
When a caller issues a call, an automatic \emph{answering machine} $M$ routes 
  the call to an available operator who then processes the call.
If $M$ receives more calls than there are available operators, 
  then $M$ routes as many calls as it can, and notifies the remaining callers
  that all operators are busy.

While the above seems reasonable, the call center in question 
  is greedy and wishes to oversubscribe its resources by contracting with a 
  second organization---thereby violating its exclusivity agreement.
This poses two problems for the call center.
First, $M$ cannot determine to which organization a call belongs; only an
  operator is in a position of making that distinction.
Second, with the current decision logic of $M$ (route to available operators,
  notify remaining callers that operators are busy), a group of callers from 
  the same organization can collectively determine that they are not being given 
  exclusive access to the call center (e.g., by placing $k$ calls 
  and noticing that not all are picked up).
Given these issues and the limit of $k$ operators (which is publicly known), 
  can the call center do anything to maintain the illusion of exclusivity?

The first observation that the call center's CEO makes is that while there 
  are $k$ operators, there is no guarantee that all of them are available at 
  any given point in time.
After all, operators are human and take breaks.
This, the CEO believes, opens the door to some level of plausible deniability.
In particular, if $M$ gives a caller from organization $O_1$ a busy signal it 
  could mean: 
\begin{myenumerate}
\item All $k$ operators are busy handling other callers from $O_1$.
\item Some operators are busy handling callers from $O_1$ and the 
  remaining operators are on a break.
\item Some operators are busy handling callers from $O_1$, some are busy
  handling callers from $O_2$, and some are on a break.
\end{myenumerate}

Possibility 1 is the expected scenario of a high-efficiency trustworthy
  call center.
Possibility 2 is an unwanted outcome since it is inefficient, but
  it does not violate the contractual agreement.
Possibility 3, however, violates the promise of exclusivity.
The goal of the call center is to design $M$ such that it is 
  hard for either of the two organizations and their callers (assume no
  coordination between organizations) to infer that possibility 3 is the one 
  taking place.
As we alluded to earlier, the key challenge is that $M$ cannot distinguish 
  between callers (and determine to which organization they belong), and 
  therefore cannot selectively lie to keep a consistent set of responses.
We thus ask whether there exists an $M$ that can leverage the proposed
  ambiguity to fool the organizations into thinking they are exclusive.
%In other words, is there a \emph{private answering machine} $M$?

We think of $M$ as acting in rounds, where in each round, $M$ receives
  a set of calls $C$.
We seek two informal properties from $M$.
\begin{myitemize}
\item \textbf{Liveness}: eventually a caller in $C$ gets to 
  talk to an operator.
\item \textbf{Privacy}: it is computationally hard for any colluding subset of 
  callers $S \subseteq C$ (some of whom may get to speak to operators) to 
  distinguish between a scenario where $S = C$ and a scenario 
  where $S \subset C$ (i.e., it is difficult for the colluding subset of 
  callers to determine whether they are the only callers or not).
\end{myitemize}

The liveness guarantee is needed for $M$ to be useful, but also 
  to rule out a trivial solution: if $M$ never puts anyone through to an
  operator, then the probability that any colluding set of callers $S$ can
  distinguish between $S = C$ and $S \subset C$ is 1/2.

\paragraph{Security game.}
To define privacy and liveness more formally, we use a security game played
  between an adversary $\Att$ and a challenger parameterized by
  a polynomial time answering machine $M$ and a security parameter $\lambda$.
$M$ takes as input a subset of callers $C$ from the set of all possible callers
  $\mathbb{C}$, a communication capacity $k$, and a random string $r$, where 
  $k = poly(\lambda), |\mathbb{C}| = poly(\lambda), |r| = poly(\lambda)$.
$M$ outputs a set of callers $U \subseteq C$, such that $|U| \leq k$.

\begin{enumerate}
\item $\Att$ is given oracle access to $M$, and can issue a 
  $poly(\lambda)$ number of queries to $M$ with arbitrary inputs $C$, $k$, $r$.
For each query, $\Att$ can observe the corresponding result 
  $U \leftarrow M(C, k, r)$. 

\item Challenger samples a random bit $b$ uniformly in $\{0, 1\}$,
  and a random string $r$ uniformly in ${\{0, 1\}}^\lambda$.

\item $\Att$ picks a set of callers $S$ (where $S \subset \mathbb{C}$) and 
  positive integer $k$, and sends them to the challenger.

\item Challenger sets $C = S$ if $b = 0$, and $C = S \cup \{e\}$ if 
  $b = 1$ (where $e$ is a uniform random element from the set $\mathbb{C} - S$).

\item Challenger calls $M(C, k, r)$ to obtain $U \subseteq C$ where $|U| \leq k$.

\item Finally, the challenger removes $e$ from $U$ (if it is present) and returns 
  the result ($U - \{e\}$) to $\Att$.\label{game:remove}

\item $\Att$ outputs its guess $b'$, and wins the security game if $b = b'$.
\end{enumerate}

In summary, the adversary's goal in the game is to determine if 
  the challenger is communicating with the uncompromised caller $e$ after 
  compromising all of the other callers (represented by $S$).

\definition[Private answering machine]{An answering machine $M$ guarantees 
  privacy}\label{def:private-machine}
  if in the above security game with parameter $\lambda$, 
  for all PPT algorithms $\Att$, there exists a
  \emph{negligible function}%
\footnote{A function $f: \mathbb{N}\rightarrow\mathbb{R}$ is negligible if
    there exists an integer $c$ such that for all positive polynomials $poly$
    and all $x$ greater than $c$, $|f(x)| < 1/poly(x)$.}
  $\textrm{negl}$ such that: 
  $|\Pr[b = b'] - 1/2| \leq \textrm{negl}(\lambda)$, where
  the probability is over the random coins of $M$ and the challenger.

\definition[Live answering machine]{An answering 
  machine $M$ guarantees liveness}\label{def:liveness} if given security
  parameter $\lambda$, for any set of callers $C$, positive 
  communication capacity $k$, and random 
  string $r$, the probability that $M(C, k, r)$ outputs a non-empty set is non-negligible in $\lambda$. 
Here $|C| = poly(\lambda), k = poly(\lambda), |r| = poly(\lambda)$,
  and the probability is over the random coins of $M$.

%% file: solutions.tex
\section{Building answering machines}%
\label{s:solutions}

We discuss two straw man proposals to highlight the challenge of
  constructing an answering machine that meets 
  Definitions~\ref{def:private-machine} and~\ref{def:liveness}.

\paragraph{Straw man $M_1$:} 
\begin{myitemize}
\item \textbf{Input}: $C, k, r$
\item $\pi \leftarrow$ uniform pseudorandom permutation of $C$ according to $r$
\item \textbf{Output}: the first $\min(k, |C|)$ elements from $\pi$
\end{myitemize}
This is not secure.
Let $X$ be the random variable describing the cardinality of the set returned
  to $\Att$, namely $|U - \{e\}|$.
Assuming that $k \leq |C|$, $\Pr[ X < k \,|\, b = 0 ] = 0$ and 
  $\Pr[ X < k \,| \,b = 1 ] = k / |C|$.
As a result, $\Att$ can, by simply counting the elements in $U - \{e\}$,
  distinguish between $b=0$ and $b=1$ with non-negligible advantage.

\paragraph{Straw man $M_2$:}
\begin{myitemize}
\item \textbf{Input}: $C, k, r$
\item $\pi \leftarrow$ uniform pseudorandom permutation of $C$ according to $r$
\item Sample $m \in [0, \min(k, |C|)]$ uniformly at random
\item \textbf{Output}: the first $m$ elements from $\pi$
\end{myitemize}

This is also not secure. Let $k = 1$ and $|S| = 1$.
The probability that the challenger returns to $\Att$ 
  an empty set is higher when $b=1$ due to
  Line~\ref{game:remove} in the security game and the way we construct $M_2$.
Again, let $X$ be the random variable describing the cardinality of the set 
  returned to $\Att$.
In particular, $\Pr[X < 1 \, | \, b = 0] = 1/2$, whereas 
  $\Pr[ X < 1 \, | \, b = 1] = 3/4$.
As a result, $\Att$ can distinguish between $b=0$ and $b=1$ with non-negligible
  advantage.
More generally, since $X$ is drawn from a uniform distribution when $b=0$, 
  the probability mass function (pmf) for $X$ (assuming $k \leq |C|$) is: 

\begin{equation*}
\begin{split}
  f(x) & = \begin{cases}
    \frac{1}{k+1} & \mbox{for  $0 \leq x \leq k$} \\
    0 & \mbox{otherwise}
    \end{cases}
\end{split}
\end{equation*}

On the other hand, if $b=1$, the pmf for $X$ is:

\begin{equation*}
\begin{split}
  f(x) & = \begin{cases}
    \frac{1}{k+1} + \frac{1}{2(k+1)} & \mbox{for $x = 0$}\\
  \frac{1}{k+1} & \mbox{for $1 \leq x \leq k-1$} \\
  \frac{1}{2(k+1)} & \mbox{for $x = k$}\\
    0 & \mbox{otherwise}
    \end{cases}
\end{split}
\end{equation*}

An adversary $\Att$ can leverage the difference in these pmfs 
  to distinguish between $b=0$ and $b=1$ with non-negligible advantage.

We could sample $m$ and permute $C$ non-uniformly, but the effect of 
  Line~\ref{game:remove} is large enough for $\Att$'s advantage to remain 
  non-negligible ($M$ must output a non-empty set with non-negligible 
  probability to satisfy liveness).
As a result, building an $M$ that guarantees privacy and liveness without
  a bound on the cardinality of $\mathbb{C}$ seems hard.
Below we give a construction under a relaxed setting.

\subsection{Machine with a bound set of callers}%
\label{s:solutions:bounded}

We now discuss the construction of an answering machine that provides privacy 
  and liveness under the assumption that the there is fixed upper bound on
  the number of possible callers ($|\mathbb{C}|$) and this bound is known 
  in advance to $M$ (the machine still does not know which callers belong to a 
  particular organization).
As a result, we assume that each element $e$ in $\mathbb{C}$ can be uniquely 
  mapped to an integer in the range $[1, |\mathbb{C}|]$ with the map 
  $id(e)$, and that this mapping is known to $M$.
The $id$ map can be set arbitrarily if the identity of all
  potential callers is known ahead of time, or populated dynamically as calls are 
  processed (a new caller $e$ is assigned a randomly unused integer in 
  $[1, |\mathbb{C}|]$ and this value is returned every time that $e$ calls). 
 
\paragraph{Private and live answering machine $M_3$:}
\begin{myitemize}
\item \textbf{Input}: $C, k, r$
\item $U \leftarrow \varnothing$
\item $\forall_{e \in C}, \forall_{0 \leq i < k}$, if $id(e) \equiv (r + i) \mod{|\mathbb{C}|}$, add
  $e$ to $U$
\item \textbf{Output}: $U$
\end{myitemize}

In other words, $M_3$ precomputes a schedule mapping callers to rounds: in 
  each round a set of $k$ callers will be serviced (the input $r$ is 
  the current round).
If a caller happens to call during a round that has been allocated for it,
  it will be added to the set $U$ (i.e., its call will be handled).
Otherwise, the call will not be answered.

Machine $M_3$ guarantees liveness because for every caller $e$, every $k$ out of 
  $|\mathbb{C}|$ rounds are assigned to $e$; since 
  $|\mathbb{C}| = poly(\lambda)$, this occurs with non-negligible probability.
Machine $M_3$ guarantees privacy because the response given to $\Att$ 
  at the end of the game (Step~\ref{game:remove} in the security game) depends 
  only on $r$ and not on $b$.
As a result this response is exactly the same when $b = 0$ and $b=1$; observing 
  this response gives no advantage to $\Att$.

%% file: attack.tex
\section{Compromised friend attack}%
\label{s:attack}

The exclusive call center problem is the scenario encountered by users 
  in MPM systems who communicate with compromised friends.
Clients in these systems can only handle a fixed number of 
  concurrent conversations in one round (this maps to the $k$ operators in 
  the call center problem), which opens the door to an attack
  that we call the \emph{compromised friend} (CF) attack.
An adversary---via compromised friends---can dial (or start a conversation 
  through any other means supported by the MPM system) a 
  client and observe whether the client responds or not.
If the client does not have a private answering machine, 
  the adversary can distinguish between a scenario where
  the client is talking to some honest client (i.e., the adversary's 
  subset of callers is not the full set, $S \subset C$), and a scenario
  where the client is not ($S = C$).
This can leak one bit of information that opens the door to existing attacks.

\paragraph{Intersection, disclosure, and hitting set attacks.}
There is a large literature of traffic analysis attacks that uncover
  patterns of communication by observing when users send and receive messages.
For example, intersection attacks~\cite{raymond00traffic} can be used to 
  narrow down the possible recipients of a message when users communicate
  with a single friend, while disclosure~\cite{agrawal03disclosure} and 
  hitting set~\cite{kesdogan04hitting} attacks can handle the case where 
  users communicate with multiple friends.
There are also statistical variants of these attacks~\cite{danezis03statistical}.

MPM systems purportedly avoid these attacks by requiring clients to always
  be online, continuously sending and retrieving messages; the client sends 
  dummy requests if the user is idle.
Unfortunately, the CF attack allows an adversary to guess whether a client 
  is sending dummy messages or not with non-negligible advantage.
An adversary can therefore target a set of potential senders and
  recipients with a CF attack, making these systems vulnerable to
  traffic analysis.
Note that the CF attack can be achieved in another way:
If a pair of friends is currently communicating at a rate
  of $r$ messages per round ($r < k$), and they wish to increase this
  rate to improve their throughput, this is the moral equivalent of 
  dialing (since it consumes a client's limited communication capacity).

\paragraph{Difficulty of conducting a CF attack in practice.}
There are some challenges in composing a CF attack with existing attacks.
First, depending on the answering mechanism of the MPM system, an adversary 
  might need to conduct a CF attack many times before learning anything useful 
  (recall from Section~\ref{s:problem} that while the adversary has 
  non-negligible advantage, it might still be small).
However, existing MPM systems~(e.g.,~\cite{angel16unobservable,
  vandenhoof15vuvuzela, lazar16alpenhorn}) currently implement an answering 
  mechanism that leaks information with a single CF attack.
Second, the CF attack requires actively targeting users on a given round, 
  which may limit the number of observations that are available to an adversary.
Last, this attack requires compromising users' actual friends 
  or it requires the use of phishing attacks to fool users into befriending 
  malicious users.

%% file: defense.tex
\section{Mitigation strategies}\label{s:defense}

We now discuss ways in which MPM systems could prevent a CF attack.
One option is for clients to use a private answering machine such as $M_3$ 
  (\S\ref{s:solutions:bounded}) to determine which of the new (or existing) 
  conversations to accept (or continue) without leaking information.
Clients would continue to exchange $k$ messages per round, but only a subset 
  of these messages (based on the output of $M_3$) would correspond to 
  actual conversations; the rest would act as cover traffic.
Note that with $M_3$, a compromised friend can learn how many other
  friends a user has, or at least an upper bound on it (i.e., $|\mathbb{C}|$).
Furthermore, $M_3$ accepts messages from a particular friend for a
  sliding window of $k$ rounds, so it is possible for two users' sliding windows 
  to never intercept.
As a result, if the MPM system does not allow the retrieval of messages
  from previous rounds, clients would be unable to communication without
  additional mechanisms.

In principle, when using $M_3$, we could set $\mathbb{C}$ to be all users in 
  the system rather than just a client's friends (so the adversary learns no 
  information); $M_3$'s function $id$ could be computed with a 
  collision-resistant hash function.
Provided that the number of total users ($n$) is $poly(\lambda$), this would 
  technically meet our liveness requirement (Definition~\ref{def:liveness}). 
In practice, however, this would result in a client accepting a call from a 
  given friend every $k$ out of $n$ rounds, which is a prohibitive delay 
  when $n$ is large.

The alternative to using a private answering machine is for clients to 
  set their communication capacity ($k$) to a value larger than their 
  maximum number of friends (under the assumption that each pair of friends 
  exchanges at most one message per round).
This too would leak the bound on the number of friends of a given client.
If a client wishes to keep this information private, a client could set $k$ to 
  be the total number of users in the system.
While this would leak no information, the communication and computational 
  costs of existing MPM systems increase linearly with $k$ (though some 
  systems have sublinear computational costs~\cite{angel16unobservable}), 
  making it prohibitive for systems with many users.
More worryingly, several MPM systems~\cite{tyagi17stadium, vandenhoof15vuvuzela} 
  provide guarantees that are based on differential privacy, and increasing
  the number of concurrent conversations ($k$) accelerates the consumption of 
  users' privacy budgets.

%% file: disc.tex
\section{Summary and discussion}\label{s:disc}

In this work we introduced an attack that allows an adversary to violate the 
  guarantees of MPM systems by leveraging users' friends.
We also proposed several mitigations, but our proposals satisfy only two out 
  of three desirable properties: privacy (leak no information), low 
  communication overhead (i.e., clients need not send many messages per round), 
  and low latency (friends get to talk to each other often).
The most pragmatic of our solutions requires bounding the maximum number of 
  friends that a client can have.

Even with our mitigations, compromised friends are a liability and can be
  used to learn sensitive information through other means.
For example, if a user is uncharacteristically slow to respond to a 
  compromised friend's message (a user's response pattern could be constructed over
  many prior interactions), this anomaly in itself leaks information.
We believe that understanding the impact of this type of attack in
  practice is a promising avenue for future work.

%% file: acks.tex
\section*{Update (10/23/2018)}
Zachary Ratliff pointed out an ambiguity in the requirement of machine $M_3$
  and the construction of the $id$ map in the version of this paper that 
  appeared at WPES.
We have clarified the requirement in this draft.

\section*{Acknowledgements}
We thank Brent Waters for a discussion that inspired this work.
We also thank Joe Bonneau, Jing Leng, Srinath Setty, Riad Wahby, Michael
  Walfish, Minjie Wang, Nickolai Zeldovich, and the WPES 2018 reviewers for
  helpful comments that improved this paper.
This work was funded by NSF grants 1514422 and 1423249; AFOSR research grant
  FA9550-15-1-0302; and ONR N00014-16-1-2154.